\documentstyle[epsf,sprocl]{article}
\newcommand{\half}{\textstyle{\frac{1}{2}}}
\newcommand{\quarter}{\textstyle{\frac{1}{4}}}

\newcommand{\beq}{\begin{equation}}
\newcommand{\beqar}{\begin{eqnarray}}
\newcommand{\eeq}[1]{\label{#1} \end{equation}}
\newcommand{\eeqar}[1]{\label{#1} \end{eqnarray}}
\newcommand{\dm}{{\mbox d}}
\def\prd#1{{Phys. Rev. }{D#1}}
\def\pre#1{{Phys. Rev. }{E#1}}
\def\prl#1{{Phys. Rev. Lett. }{#1}}
\def\pl#1{{Phys. Lett. }{#1}}
\def\np#1{{Nucl. Phys. }{#1}}
\def\ap#1{{Ann. Phys. (N.Y.) }{#1}}
\def\prep#1{{Phys. Rep. }{#1}}

\def\rmp#1{{Rev. Mod. Phys. }{#1}}

\begin{document}

\title{OPEN QUANTUM SYSTEMS, ENTROPY AND CHAOS\,\footnote{
Invited talk presented at the 5th Rio de Janeiro International Workshop 
on Relativistic Aspects of Nuclear Physics,  
August 1997 --- 
to be published in the proceedings, eds. 
T. Kodama et al. (World Scientific).}}

\author{HANS-THOMAS ELZE}

\address{Universidade Federal do Rio de Janeiro, Instituto de F\'{\i}sica \\
Caixa Postal 68.528, 21945-970 Rio de Janeiro, RJ, Brasil \\ 
and \\  
Physics Department, University of Arizona, Tucson, AZ\,85721}

\maketitle\abstracts{Entropy generation in quantum sytems 
is tied to 
the existence of a nonclassical environment (heat bath or other)
with which the system interacts. The continuous `measuring'
of the open system by its environment induces decoherence 
of its 
wave function and entropy increase. Examples of nonrelativistic 
quantum Brownian motion and of interacting scalar fields 
illustrate these general concepts. It is shown 
that the Hartree-Fock approximation 
around the bare ($\hbar =0$) classical limit can lead to 
spurious semiquantum chaos, which may affect the 
determination of entropy production and thermalization also 
in other cases.}

\section{Introduction}
The related topics of environment induced quantum 
decoherence and the transition from quantum to 
classical mechanics have recently been investigated 
in widely varying contexts, ranging from  
problems of interpretation of quantum mechanics, of the 
measurement process in particular, to questions of 
how classical physical laws and the observed classical features of cosmology emerge in the underlying quantum universe, see e.g. 
Refs.\,\cite{Zu}--\,\cite{Ellis}. 
The central idea of the quantum decoherence approach is 
that the transition {\it quantum} $\rightarrow$ 
{\it classical} can be understood as a dynamical effect  within quantum mechanics itself, especially keeping $\hbar$    
nonzero as it is.   
  
In this lecture I will explain some of these 
issues, in particular how {\it entropy generation} is intimately related to an {\it open quantum  
system}. Its identification, its 
properties, and the calculation of the time dependent 
entropy production present unresolved problems for  
strongly interacting systems, such as the high 
energy density matter formed during relativistic  
hadronic or nuclear collisions. Based on more technical details given in Refs.\,\cite{I}, 
I will outline how the quantum decoherence approach can 
be applied here. 
    
This is obviously related to the persistent issue of thermalization 
of strongly interacting matter, the assumption of which 
is one of the conceptual cornerstones guiding the ongoing search for 
the quark-gluon plasma (QGP), see Ref.\,\cite{QM} and earlier ``Quark Matter'' proceedings. As reflected during 
this conference, many      
theoretical attempts to pinpoint the most relevant QGP properties depart from the assumption of a thermalized 
high entropy density state of matter. However, at present 
we are still quite far from understanding 
how an initially `pure' quantum state with zero entropy,   
e.g. two colliding nuclear wave packets,      
could evolve into the `mixed' QGP state.   
  
In order to fully appreciate the problem, we need a 
precise definition of the entropy, which preferably 
should allow us to extrapolate to the usual thermodynamical 
limit or Boltzmann's entropy of statistical mechanics.  
Let us recall that  
the First Law of Thermodynamics relates infinitesimal 
changes of the internal energy $U$, the volume $V$, and 
the entropy $S$, given the pressure $P$ and the temperature 
$T$ of the system: 
\beq 
\dm U=-P\dm V+\delta Q=-P\dm V+T\dm S
\;\;. \eeq{firstlaw}
Thus, the change of the internal energy is related to the 
work done by the system against the external pressure 
and to the heat $\delta Q$ transfer or the entropy change, 
respectively. The Second Law of Thermodynamics rules 
that the entropy in a {\it closed} system cannot decrease:
\beq 
\frac{\dm S(t)}{\dm t}\ge 0
\;\;. \eeq{secondlaw} 
Furthermore, based on a statistical definition  
of the ($N$-particle) entropy,  
\beq
S_B(t)\equiv -k_B\int
\frac{\dm^{3N}x\;\dm^{3N}p}{(2\pi\hbar )^{3N}}\;
f(x,p;t)\;\mbox{ln}f(x,p;t)
\;\;, \eeq{SB}
employing a suitable phase-space distribution function, 
one is led to the interpretation due to Boltzmann that 
the entropy measures the observer's {\it lack of information} 
about the system. It determines the number of 
possible realizations (microstates) of the system which 
conform with a given macrostate specified, for 
example, in terms 
of the above thermodynamical variables. 
  
Evidently, in relativistic heavy-ion 
physics much use is made of such related 
thermodynamical notions as equilibration, 
thermal $p_\perp$-spectra, equation of state, 
phase transition, etc.   
  
It is our aim here to get a quantum mechanical handle 
on the entropy. -- As is well known, the Wigner function 
is what comes closest 
to a quantum mechanical distribution function replacing 
$f(x,p;t)$ \cite{Feyn}; see Ref.\,\cite{EH} for a review of earlier  
work on quark-gluon transport theory, i.e. the dynamics 
of QCD Wigner functions. For our present purposes it is  
more convenient to work with density matrices, which 
generally are related to Wigner functions by appropriate 
Fourier transforms. -- In terms of the 
density operator $\hat\rho$ characterizing a (closed) 
quantum mechanical system, we will henceforth employ 
the von\,Neumann definition of the entropy \cite{Zu,Feyn}: 
\beq
S(t)\equiv -\mbox{Tr}\;\hat\rho (t)\;\mbox{ln}\hat\rho (t)
=-\sum_n\;w_n\;\mbox{ln}w_n 
\;\;\, \eeq{S}
where the trace is over a complete set of states and 
$w_n$ denotes the probability of such a state $|n\rangle$\,; 
from now on we employ units such that $\hbar =c=
k_B=1$\,, except when stated otherwise. 

As we 
shall see in the next section, the von\,Neumann 
entropy has several desirable properties, but shows some 
surprising features as well. Most notably 
$S\equiv 0$ for closed systems. The latter result 
has led to some confusion in the past \cite{EC}, in particular it 
has been claimed that $S$, as defined here, is completely 
useless, if one wants to characterize the entropy 
production in high-energy physics. However, 
these objections are overcome by the decoherence 
approach and especially by realizing that in this case, like in   
most if not all physically interesting cases, the system is open indeed.   

\section{The von\,Neumann Entropy and the Need for Open Systems}
  
In order to illustrate some important features of the entropy $S$, as defined in eq.\,(\ref{S}), we evaluate 
it for two limiting cases, for a quantum system in a pure 
state and in the thermal equilibrium state, respectively. 
An elementary example for a two-state system is given 
in Ref.\,\cite{I}(c). 
 
If the system is in a {\it pure state} $|\Psi\rangle$, the 
density operator assumes the simple form 
$\hat\rho =|\Psi\rangle\langle\Psi |$\,. Correspondingly, the probability of finding the system in this state is 
$w_{\Psi}=1$. Therefore, we obtain immediately: 
\beq
S_{\Psi}=-1\cdot\mbox{ln}1=0 
\;\;, \eeq{Spsi} 
which can be interpreted that we know everything about the 
system that we possibly can. 

Similarly one finds that 
if the system is in any one of $N$ states with equal 
probability $1/N$, then the entropy equals ln$N$\,, i.e.  
the maximum value, when we are completely 
ignorant about which state the system is in.  
  
Let us consider one of the most important cases of   
an impure or {\it mixed state}, namely when the system 
is in thermal equilibrium. Then, the density operator 
is given by: 
\beq
\hat\rho =Z^{-1}{\mbox e}^{-\beta\hat H}\;\;,\;\;\;
Z\equiv\mbox{Tr}\;{\mbox e}^{-\beta\hat H}=\sum_n\;
{\mbox e}^{-\beta E_n}
\;\;, \eeq{rhotherm} 
where $Z$ is the partition function, $\hat H$ is the 
Hamiltonian, and $\beta\equiv
T^{-1}$ denotes the inverse temperature of the system. 
We find: 
\beqar
S(T)&=&-\sum_n\;\frac{{\mbox e}^{-\beta E_n}}{Z}\;
\mbox{ln}\frac{{\mbox e}^{-\beta E_n}}{Z}
\; =\;\mbox{ln}Z-\mbox{Tr}\;\frac{{\mbox e}^{-\beta\hat H}}{Z}
(-\beta\hat H)
\nonumber \\[2ex]
&=&\mbox{ln}Z-Z^{-1}\beta\partial_\beta Z\; =\;\partial_T
(T\mbox{ln}Z)\;\equiv\; -\partial_TF
\;\;, \eeqar{ST} 
which relates the von\,Neumann entropy to the usual 
partial derivative of the free energy, which is familiar 
from thermodynamics. 
  
The properties of the entropy which we have considered 
so far tie in with our knowledge of thermodynamics 
or classical statistical mechanics. However, a problem 
arises immediately, since the entropy $S$ turns out to 
be a constant of motion. According to the Schr\"odinger 
equation the time evolution of wave functions is unitary, 
$|\Phi (t)\rangle =\mbox{exp}(-i\hat Ht)|\Phi (0)\rangle$\,, which implies:  
\beq
S(t)=-\mbox{Tr}\left\{ {\mbox e}^{-i\hat Ht}\hat\rho (0)
{\mbox e}^{i\hat Ht}\mbox{ln}[{\mbox e}^{-i\hat Ht}     
\hat\rho (0){\mbox e}^{i\hat Ht}\right\}=S(0)
\;\;, \eeq{Sconst}
using the cyclic property of the trace.     
In a closed system the entropy stays constant at the  
initial value. 
  
More specifically, the unitary quantum mechanical time 
evolution prohibits a transition from a zero entropy 
pure initial state to a mixed nonzero entropy final state. 
However, it is well known how to overcome this impasse.         
Similarly as in classical statistical mechanics, where 
coarse graining in phase space allows to circumvent 
$\dm f/\dm t=0$ (Liouville flow) and hence $\dot S_B=0$, cf. eq.\,(\ref{SB}), we have to give up the 
premise of unitary time evolution.  

The decoherence approach is based on a separation of 
`all degrees of freedom' into the observed degrees of 
freedom ({\it system}) and the `rest of the universe' 
({\it environment}) \cite{Zu,Om,GH}. 
The border between system 
and environment generally has to be open to the exchange of 
information and possibly to the exchange 
of energy, momentum, etc. 
Without interaction 
two isolated closed systems result.  
Then, the density 
operator $\hat\rho$ representing all degrees of freedom (d.o.f.) 
evolves unitarily and the corresponding entropy is a  
constant, as we have shown. However, the density operator 
$\hat\rho_S$ representing the system, 
\beq 
\hat\rho_S(t)=\mbox{Tr}_E\hat\rho (t)
\;\;, \eeq{rhosys} 
which is obtained by tracing over the environment degrees 
of freedom, evolves in a more complicated non-unitary way. 
Consequently, the entropy of the system $S_S(t)$\,, which is defined as the von\,Neumann 
entropy employing $\hat\rho_S(t)$\,, will change in time,
as we shall see. 
  
The non-unitary 
evolution of the system density matrix is accompanied 
by {\it environment induced decoherence}, i.e. typically 
its off-diagonal elements which encode the quantum 
mechanical interference effects decay with a characteristic 
decay constant $\tau_D$\,. Correspondingly, the eigenvalues 
of $\hat\rho_S$ change and the diagonal matrix elements 
become probabilities characterizing the 
resulting mixed state. If there is thermalization, 
\beq
\rho_S^{nn}(t)\stackrel{\tau_{th}}{\longrightarrow} 
\frac{{\mbox e}^{-\beta E_n}}{Z}
\;\;, \eeq{thermal} 
i.e. the diagonal matrix elements become the usual  
Boltzmann weights, cf. eqs.\,(\ref{rhotherm}),\,(\ref{ST}). Most interestingly, 
the decoherence time $\tau_D$ is found to be several ten orders 
of magnitude smaller than the thermalization time 
$\tau_{th}$ for models of macroscopic bodies interacting 
with their environment (cosmic background radiation, atmosphaeric gas particles, etc.)\cite{Te}. This 
explains the classical behavior of most of what we 
observe in daily life \cite{Zu}. Cases of 
quantum decoherence and `revival' of the off-diagonal matrix elements are also known and experimentally observed in cavity quantum electrodynamics.  
 
For microscopic systems, generally, the situation 
is more intricate and depends sensitively on the 
interactions and the  
separation into system and environment d.o.f.,  
in particular for more complex systems 
than a single particle or one d.o.f. interacting with an environment \cite{Te}. 
If we find $\tau_D\ll\tau_{th}$\,, then 
the possibility arises of creating a system with high 
entropy which is not at all thermalized. We would like  
to know precisely how these time scales compare in 
high-energy collisions of hadrons or nuclei and, of course,
to what extent these systems thermalize.
  
Next, we present the useful 
theorem that the entropy 
generated in a system exactly equals the entropy 
generated in its environment due to the {\it mutual} decoherence process, 
\beq
S_S(t)=-\mbox{Tr}\;\hat\rho_S(t)\;\mbox{ln}\hat\rho_S(t)
=-\mbox{Tr}\;\hat\rho_E(t)\;\mbox{ln}\hat\rho_E(t)
=S_E(t)
\;\;, \eeq{Sequal}
with $\hat\rho_E$ calculated from 
$\hat\rho$ by tracing over the system d.o.f. 
analogously to $\hat\rho_S$\,, eq.\,(\ref{rhosys}). 
This result follows 
from the Schmidt decomposition of the 
density matrices \cite{I}(a), 
which is for matrices (e.g. in Hilbert space) what the 
Schmidt orthogonalization procedure is for vectors. 
    
This leads us to the question, whether 
the hard as well as the large number of soft 
{\it photons}     
generated by bremsstrahlung from quarks during the initial 
hard scattering and stopping phase of a hadronic collision 
yield an important decoherence effect. This would 
imply that a major part of the entropy observed in the 
final hadronic states of the reaction is already generated 
during its initiation. A simple counting argument based 
on the ratio of effective photon and QGP 
d.o.f., i.e. their entropy ratio in thermal equilibrium,
\beq 
\frac{N_\gamma}{N_{QGP}}=\frac{2\tilde\sigma VT^3}
{37\tilde\sigma VT^3}
\;\;, \eeq{Nratio} 
seems to indicate that only about 5\% (or less) of the 
entropy can be generated in this way. Note, however, the 
strong temperature dependence of the absolute numbers. 
Even if it does not maximize the entropy, as a thermal 
distribution does, a nonequilibrium distribution generated by bremsstrahlung processes which effectively   
corresponds to a higher photon `temperature' may 
lead to a sizeable correction of the above estimate. The discussion   
following eq.\,(\ref{Spsi}) explains that generally a
flat distribution contains less information/more entropy than a steep 
one and that the size of the available phase space matters (`ln$N$').    
A dynamical calculation generalizing  
results presented in the following Sections 3 and 4 is presently carried out.   
     
In the example  
just considered the separation between system and environment d.o.f. is based on the fact that the   
strongly interacting system of quarks and gluons is essentially bound    
for, say, 10\,fm/c until `freeze-out', whereas the 
nonequilibrium  
environment photons are generated particularly during the first 0.1\,...\,0.5\,fm/c and essentially free to leave.
Thus the `rest of the universe' or {\it environment} 
is open to interpretation 
and has to be determined according to the physical circumstances.    
 
\section{A Nonrelativistic Quark in Brownian Motion}  
 
The purpose of this section is to demonstrate how the general 
concepts of environment induced decoherence and entropy generation 
in an open quantum system work in practice. The example can be understood 
as a nonrelativistic (heavy) quark moving and interacting with its 
own gluonic environment field, the properties of which can only be  
modelled at present.  
  
We start with a model Hamiltonian describing the bilinear 
translation invariant interaction between a nonrelativistic particle 
({\it system}) and an infinite set of harmonic oscillators ({\it environment}):  
\beq 
H=\frac{p^2}{2M}+\sum_{\omega_n\leq\Omega}\left\{
\frac{P_n^2}{2m_n}+\half m_n\omega_n^2(X_n-x)^2\right\}
\;\;, \eeq{H} 
where $\Omega$ presents a high-frequency cut-off (such as Debye frequency 
and inverse classical electron radius in similar models related to the 
polaron and electron, respectively); we may 
think of the QCD scale 
parameter $\Lambda_{QCD}\approx 200$\,MeV\,. This type of 
model associated with the names of Feynman and Vernon or Caldeira and 
Leggett has been extensively studied in the field of quantum 
Brownian motion \cite{Grab}, developing further the original
Feynman-Vernon influence functional technique.    
  
As it turns out, the physics described by the model of eq.\,(\ref{H}) 
is completely determined by the spectral density distribution of the 
environment:   
\beq
I(\omega)\equiv\half\sum_{\omega_n\leq\Omega} m_n\omega_n^3
\delta (\omega -\omega_n)
\;\;, \eeq{sdd}
and the related noise and dissipation kernels \cite{Grab}. 
-- For the polaron and electron cases one obtains $I(\omega)\propto\omega^3$ 
for $\omega\rightarrow 0$\,, which lead to the well known 
mass renormalization for $t\rightarrow\infty$\,, whereas for an `Ohmic' environment one chooses $I(\omega)\propto\omega$, leading to a dissipative force proportional 
to the velocity of the Brownian particle. 

Unfortunately, in the QCD 
case we do not know 
how to calculate the spectral density distribution. Classically 
QCD is a nonintegrable theory, which implies that such a 
decoupled oscillator representation of the dynamics does {\it not} exist
\cite{Schuster}. 
Therefore, it seems unlikely that standard perturbation theory, 
which is based on the $\hbar$-(loop-)expansion around the naive classical 
limit, can provide reliable results in this context.  

Keeping this limitation in mind, I considered this model in 
the {\it strong-coupling limit} for $t\ll\Omega^{-1}$ at $T=0$\,, 
where only certain 
moments of $I(\omega)$ enter the calculation and can be chosen as 
parameters of the model. Presently the  
spectral density is much larger in the infrared than in 
the polaron/electron case. Technical details of the calculation 
of the reduced density matrix propagator in the environment corresponding 
to eq.\,(\ref{H}) can be found in Ref.\,\cite{I}(a). Also  
longer time scales may be interesting to study.       
  
Here we want to discuss only the very initial  
influence of the environment. Choosing 
$I(\omega)\equiv g\Omega^3F(\omega /\Omega )\Theta (\Omega -\omega )$\,, 
with an undetermined shape function $F$ and dimensionless coupling 
constant $g$\,, we have $g_0\equiv g\int_0^1\dm x\;x^{-1}F(x)$\,. Then, 
the following parameter and dimensionless time variable, respectively,  
are needed: 
\beq 
\alpha\equiv\frac{1}{2g_0}\frac{\Omega}{M}(w_0\Omega )^{-4}\;\;,\;\;\;
t_+\equiv\Omega t(2g_0\frac{\Omega}{M})^{1/2}
\;\;, \eeq{params} 
where $w_0$ denotes the initial ($t=0$) width of a Gaussian wave packet 
representing a particle moving with the 
velocity $v_0=\langle p\rangle_0/M$ initially.
  
\begin{figure}[htbp]
\centerline{
\epsfysize=14.50cm
\epsffile{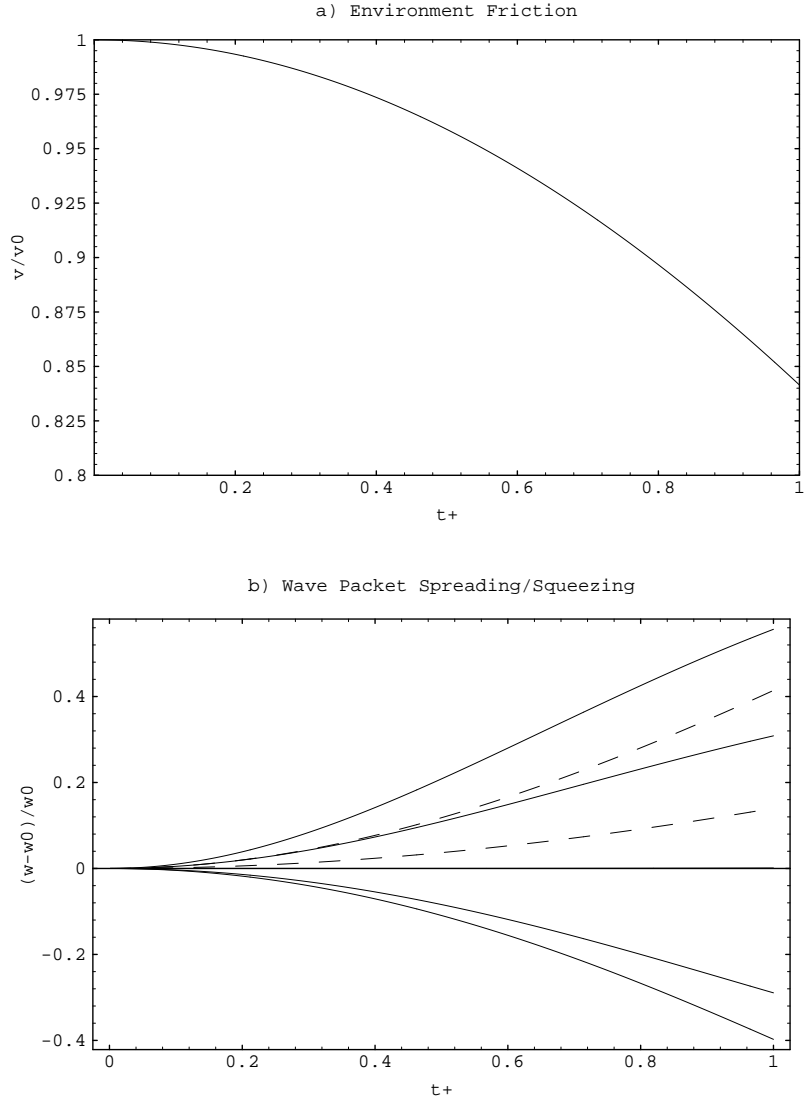}}
\caption{Environment effects in nonrelativistic quantum Brownian motion. 
a) The center of a Gaussian wave packet slows down. b) The width of the 
wave packet ($\alpha =3.0,\,2.0,\,1.0,\,0.3,\,0.1$\,, full lines, top to bottom) expands more rapidly or may be squeezed as 
compared to the usual case ($\alpha = 1.0,\,0.3$\,, upper 
and lower dashed curve, respectively). See main text for details.\hfill .}
\end{figure}
\begin{figure}[htbp]
\centerline{
\epsfysize=14.50cm
\epsffile{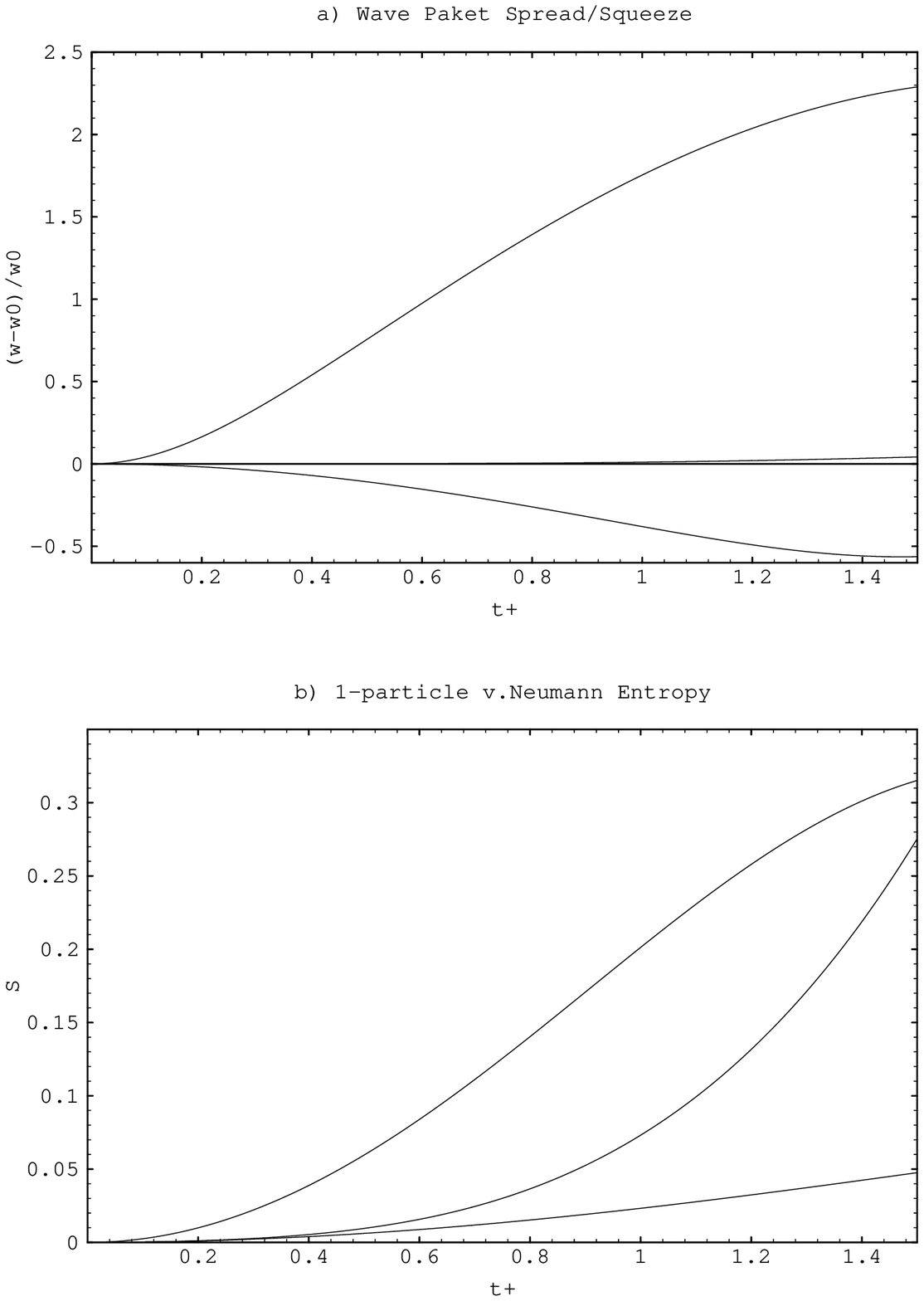}}
\caption{(a) Similar as Fig.\,2\,b)\,, however, more extreme  
initial widths are considered -- small, medium, and large wave packets 
(top to bottom), respectively. (b) The corresponding single particle 
entropy for a large, small, and medium size wave packet (top to bottom). 
See main text for details.\hfill .}
\end{figure}

In Fig.\,2\,a) we observe the dissipative `friction' effect of the environment. In the present approximation we obtain a universal curve showing 
the deceleration of the particle in terms of the time dependent 
velocity (divided by $v_0$) as a function of $t_+$\,. For $g=0$, of course, $v(t_+)=v_0$\,. 
Secondly, as shown in 
Fig.\,2\,b), we find some  
surprising effects which the environment can have on the 
width of the wavepacket. There, the relative change of the width as
a function of $t_+$ is represented: the wave packet spreads faster ($\alpha >1$) than a corresponding free ($g=0$) quantum mechanical wave packet, the width remains constant ($\alpha =1$), or it even gets squeezed ($\alpha <1$),  depending on the parameter $\alpha$\,. The dashed curves show the spreading for $g=0$; they do not coincide because of the rescaled time axis. If the other  
physical model parameters were fixed, this result clearly shows 
that the particular coordinate-coordinate coupling between 
the particle and its environment, which we have 
chosen in eq.\,(\ref{H}), is sensitive to the size of the particles' wave packet: sufficiently small packets spread 
faster than usual, sufficiently large ones get squeezed.    
  
This is illustrated once more for some more extreme cases 
in Fig.\,3\,a), where the three curves (top to bottom) 
correspond to an initially small ($w_0\Omega\ll 1$), 
medium size ($w_0\Omega\approx 1$), and large 
($w_0\Omega\gg 1$) wave packet, respectively.
For these three cases 
we show in Fig.\,3\,b) the corresponding entropy, i.e. 
the entropy evaluated with the help of the time dependent 
density matrix of the nonrelativistic particle. In this figure the curves correspond to an initially large (top), small (middle), and medium size (bottom) wave packet, respectively. We do not attribute too much 
importance to the sizeable entropy which can be reached in very short time, e.g. 0.1\,fm/c\,, 
since the numbers   
depend on the model parameters; we recall, however, the  
ultrarelativistic/high-$T$ limit with 4 units of entropy 
per particle. -- Actually 
represented here is the simpler {\it linear entropy}, 
$S_{lin}\equiv\mbox{Tr}[\hat\rho -\hat\rho^2]$\,, which  
provides a lower bound for the von\,Neumann entropy 
\cite{I}(a).  
The linear and von\,Neumann entropies 
coincide in the zero and maximal entropy limits. -- It is remarkable that medium size wave packets, the width of
which is least affected by the environment, also generate the 
lowest entropy increase. They essentially behave like a {\it classical}
particle. 
 
Furthermore, coherent superpositions of such states decohere most quickly and 
show the strongest entropy generation \cite{I}(a). These properties taken together 
qualify the Gaussian wave packets as approximate {\it pointer states} 
of the present model, similarly to the simple model 
studied in Ref.\,\cite{ZHP}. Pointer states show the behavior of 
a classical pointer \cite{Zu,Om,Ze}: coherent superpositions 
of such states decohere very rapidly into mixed states, which are   
observed as various pointer positions with certain probabilities in a (macroscopic) experiment.               

\section{Interacting Scalar Fields}  

In this section we study two interacting scalar quantum 
fields, representing the {\it system} (`1') and {\it environment} (`2'), 
respectively. We want to describe their dynamics as well 
as the entropy generated in the system. We sketch this  first step in generalizing the Brownian motion example 
of Section 3
towards more realistic cases related to high-energy 
collisions, while the technical details may be found in 
Refs.\,\cite{I}.     
  
Let the classical action of this model be given as: 
\beq 
S\equiv\int\dm^4x\left\{L_1+L_{12}+L_2\right\} 
\;\;, \eeq{actionC} 
with the interaction potential $L_{12}=-V(\Phi_1,\Phi_2)$, 
and the Lagrangians: 
\beq
L_j\equiv\half\partial_\mu\Phi_j\partial^\mu\Phi_j-v_j(\Phi_j)\;\;,\;\;\;
v(\Phi )\equiv -\frac{1}{2}\mu^2\Phi^2+\frac{1}{4!}\lambda
\Phi^4
\;\;. \eeq{L}
Consider the separation into system and environment 
as originating from splitting a single scalar field, 
$\Phi\equiv\Phi_1+\Phi_2$, according to some `coarse 
graining prescription' to be discussed shortly. Then, 
the interaction potential is: 
\beq 
V(\Phi_1,\Phi_2)=-\mu^2\Phi_1\Phi_2-\partial_\mu\Phi_1
\partial^\mu\Phi_2+\frac{1}{4!}\lambda\Phi_1\Phi_2
[4\Phi_1^{\; 2}+6\Phi_1\Phi_2+4\Phi_2^{\; 2}]
\;\;. \eeq{V} 
We see that the bilinear coupling studied in most 
quantum Brownian motion models appears naturally here
among other terms. 
Presently we will keep only this first term for 
simplicity, while results for a general quartic 
$V(\Phi_1,\Phi_2)$ are derived in Ref.\,\cite{I}(a). --   
Now, various interpretations of the  
fields $\Phi_1,\,\Phi_2$ are possible: 
\begin{itemize}
\item Coordinate space: $\Phi_1$ and $\Phi_2$ represent 
the same field, however, with support either inside the system or in the environment, respectively; 
$V(\Phi_1,\Phi_2)$ presents a contact interaction at the 
surface. Models of this type have been studied for the 
geometric entanglement entropy generated by event 
horizons (black holes), see e.g. Refs.\,\cite{Mark}. 
\item Momentum space: $\Phi_1$ and $\Phi_2$ represent 
the long- and short-wavelength components of the same 
field. This has been of much interest in cosmology in 
the context of inflationary models; Ref.\,\cite{LM} is 
a recent example, from which other references can be traced.   
\end{itemize} 

Strongly interacting matter naturally leads to a 
separation of d.o.f. 
In a finite size QGP droplet, with radius $R$ on the order 
of a few fm\,, 
quark-gluon modes with wavelengths  
much larger than $R$ are suppressed due to 
confinement and form an environment of 
virtual fluctuations together with composite meson 
(baryon) fields in the nonperturbative vacuum.   
Its properties, as with the case discussed in 
Section 3\,, cannot be calculated at present, but 
presumably play 
a crucial role in multiparticle production processes and 
the associated entropy generation. Finally, we mentioned 
before the bremsstrahlung photon environment which contributes to the decoherence of the charged quark system. 
 
In the following we assume $\Phi_1$ and $\Phi_2$ to be two physically 
distinct fields. 
 
\subsection{Variational Approach to Time Evolution in 
Field Theory}

We now turn to the dynamics of the coupled fields of 
our model defined in eqs.\,(\ref{actionC})--(\ref{V}). The  
standard approach is motivated by perturbation theory, 
i.e. the usual $\hbar$-(loop-)expansion, and based on 
the Schwinger-Keldysh formalism. In order to resum the  
large class of (iterated bubble) 
diagrams corresponding to the time dependent 
Hartree-Fock approximation (TDHF), it is much more   
efficient to employ Dirac's time dependent variational
principle. 

The starting point is the
{\it functional Schr\"odinger equation} describing the full dynamics
of a generic field $\varphi$ in the Schr\"odinger picture 
\cite{Jackiw}:
\beq
i\partial _t\Psi [\varphi;t]=
H[\hat{\pi},\varphi ]\Psi [\varphi ;
t]\equiv\int\dm^dx\left \{-\half\frac{\textstyle{\delta ^2}}
{\textstyle{\delta\varphi ^2}}+\half (\nabla\varphi )^2+{\cal V}(\varphi
)\right \}\Psi [\varphi ;t]
\;\;, \eeq{85}
where $\Psi [\varphi ;t]\equiv\langle\varphi |\Psi (t)\rangle$ denotes
the wave functional in the $\varphi$-representation, which
corresponds to a wave function $\psi (x,t)
\equiv\langle x|\psi (t)\rangle$ for a one-dimensional
quantum mechanical system, and $\hat{\pi} =-i\delta /\delta\varphi$
represents the canonical momentum operator conjugate to the field
(`coordinate') $\varphi$\,. The dynamics is determined by the
Hamiltonian $H$. In this context the completeness
and inner product
relation, respectively, involve functional integrals instead of
ordinary ones (orthogonality needs a $\delta$-functional),
for example:  
\beq
\langle\Psi _1(t)|\Psi _2(t)\rangle\equiv
\int {\cal D}\varphi\;\langle\Psi _1(t)|\varphi\rangle\langle\varphi |
\Psi _2(t)\rangle =\int {\cal D}\varphi\;\Psi _1^\ast [\varphi ;t]
\Psi _2[\varphi ;t]
\;\;, \eeq{86} 
with details to be found in Ref.\,\cite{I}(a). 

Then, we may state the
{\it variational principle}:
\beqar
\frac{\textstyle{\delta\Gamma [\Psi ]}}{\textstyle{\delta\Psi}}&=&0
\;\;,\;\;\;\mbox{for all $\Psi$ with}\;\;\;\langle\Psi (t)|\Psi (t)
\rangle =1\;\;, \label{89} \\ [2ex]
\mbox{where}\;\;\;\Gamma [\Psi ]&\equiv&\int\dm t\;\langle\Psi (t)|
[i\partial _t-H]|\Psi (t)\rangle
\;\;, \eeqar{90}
i.e. requiring the effective action $\Gamma$ defined in eq.\,(\ref{90}) to be stationary against arbitrary variations of the
normalized wave functional $\Psi$\,, which vanish at $t\rightarrow\pm
\infty$, is equivalent to the exact functional
Schr\"odinger equation, eq.\,(\ref{85}) above. With the variational
principle in hand, one can solve the
time evolution problem approximately by
choosing a suitably parametrized
trial wave functional. We remark that the effective action is real 
by construction.
 
We work with the most general Gaussian
trial wave functionals. For a generic field $\varphi$ it is defined by:
\beq
\Psi _G[\varphi ;t]\equiv N(t)\exp \{
-[\varphi -\bar{\varphi}(t)]
\left [\quarter G^{-1}(t)
-i\Sigma (t)\right ]
[\varphi -\bar{\varphi}(t)]
+i
\bar{\pi}(t)
[\varphi -\bar{\varphi}(t)] \}
\;, \eeq{91}
where henceforth we do not explicitly write the integrations over spatial variables.
For example,
\beqar
\varphi G^{-1}(t)
\bar{\varphi}(t)&\equiv&
\int\dm^dx\dm^dy\;
\varphi (x)G^{-1}(x,y,t)
\bar{\varphi}(y,t) \;\;, \nonumber \\ [1ex]
\bar{\pi}(t)
\varphi &\equiv&\int\dm^dx\;
\bar{\pi}(x,t)
\varphi (x) \;, \;\;
\mbox{tr}\;\Sigma (t)\;
\equiv\;\int\dm^dx\;\Sigma (x,x,t)
\;. \eeqar{92}
The normalization factor $N$ (for symmetric
and positive-definite $G$\,) is: 
\beq
1=\int {\cal D}\varphi\;\Psi _G^\ast [\varphi ;t]
\Psi _G[\varphi ;t]\;\;\longrightarrow\;\;\;
N(t)=({\cal N}\;\mbox{det}\; G(t))^{-1/4}
\;\;, \eeq{93}
cf. eq.\,(\ref{86}) 
(${\cal N}$ is an infinite constant which we omit henceforth). 

The meaning of the variational parameter functions $\bar{\varphi}$,
$\bar{\pi}$, $G$, and $\Sigma$ is discussed  
in Ref.\,\cite{I}(a). 
The choice of Gaussian trial wave functionals is dictated by the need to evaluate   
functional integrals, in order to calculate the effective 
action $\Gamma$\,, eq.\,(\ref{90}). 
The equivalence of this Hartree-Fock effective action with the Cornwall-Jackiw-Tomboulis
generating functional for two-particle irreducible
graphs was demonstrated in Ref.\,\cite{Corn} for
energy eigenstates of the field.
Variation of $\Gamma$ 
w.r.t. to the parameter functions will give the dynamical 
equations, cf. Section 4.3 below,  
representing the
field theory in terms of coupled equations for the 
one- and two-point Wightman
functions. 

We proceed with the ansatz for the two-field 
wave functional: 
\beqar
\Psi _{12}[\Phi _1,\Phi _2;t]&\equiv&N_{12}(t)\;\Psi _{G_1}[\Phi _1;t]\;
\Psi _{G_2}[\Phi _2;t]
\\ [1ex]
&\cdot&
\exp\left\{-\half [\Phi _1-\bar{\Phi} _1(t)]\left [G_{12}(t)-i
\Sigma _{12}(t)\right ]
[\Phi _2-\bar{\Phi} _2(t)]\right\}
\;\;, \nonumber \eeqar{100}
with the normalized Gaussians on the r.h.s. as defined in eq.\,(\ref{91}), all terms carrying suitable indices  
for the field they belong to, and where 
$N_{12}$ denotes an additional normalization
factor. The latter is necessary, since we included an essential
exponential describing the {\it two-point correlations} between the system and the 
environment fields. We obtain:    
\beq
N_{12}(t)=
\left (\det\{ 1-G_1(t)G_{12}(t)G_2(t)G_{12}(t)\}\right )^{1/4}
\;\;, \eeq{101}
similarly as in eq.\,(\ref{93}).
 
Employing eq.\,(\ref{100}), the calculation of the two-field effective action 
$\Gamma$ is straightforward, even if tedious.   
We present here only the result for the 
simplest bilinear interaction, $V(\Phi_1,\Phi_2)\equiv 
-\mu^2\Phi_1\Phi_2$\,, while the case of a general 
quartic interaction is studied in Ref.\,\cite{I}(a): 
\beqar
\Gamma [\Psi_{12}]&=&\left .\int\dm t\;\right\{
\sum_{j=1,2}\left\{\;
\bar{\Pi}_j\dot{\bar{\Phi}}_j
-\half\bar{\Pi}_j^{\; 2}
-\half (\nabla\bar{\Phi}_j)^2-v_j(\bar{\Phi}_j) +\half\mu^2\bar{\Phi}_1\bar{\Phi}_2 \right .
\nonumber \\ [1ex]
&\;&\;\;\;\;\;\;\;\;+\;\hbar\;\mbox{tr}\; [\;
\Sigma_j\dot{\bar{G}}_j
-2\tilde{\Sigma}_j^{\; 2}\bar{G}_j
-\frac{1}{8}G_j^{-1}[A+B]
+\frac{1}{2}\nabla ^2\bar{G}_j\; ]
\nonumber \\ [1ex]
&\;&\;\;\;\;\;\;\;\;-\;\frac{\textstyle{\hbar}}
{\textstyle{2!}}\langle v_j^{(2)}
\rangle\;\mbox{tr}\; \bar{G}_j
-\frac{\hbar^2}{\mbox{V}_d}
\frac{\textstyle{3}}{\textstyle{4!}}\left .v_j^{(4)}
(\;\mbox{tr}\;\bar{G}_j)^2
\;\right\}
\nonumber \\ [1ex]
&\;&\;\;\;\;\;\;\;\;+\;\left . 
\frac{\hbar}{2}\;\mbox{tr}\; [\;\dot{\Sigma}_{12}\bar{G}_1\bar{G}_2
\bar{G}_{12}\; ]
-\hbar\mu^2
\mbox{tr}\; [\bar{G}_1\bar{G}_2
\bar{G}_{12}]\right\} 
\;\;, \eeqar{124}
with $v_j^{(n)}\equiv d^nv_j(\bar{\Phi}_j)/d\bar{\Phi}_j^{\;n}$\,,
$\langle f\rangle\equiv\int\dm^dx\,f(x)/V_d$\,, and 
$V_d\equiv\int\dm^dx$\,. We also use the
abbreviations $\bar{G}_{12}\equiv G_{12}[A-B]$\,,
$\bar{G}_j\equiv [A-B]^{-1}G_j$\,, and $\nabla ^2$ acts on either
one of the two formal spatial coordinates of $\bar{G}_j$\,; 
furthermore,  
$\tilde{\Sigma}_1\equiv\Sigma_1-\quarter G_2G_{12}\Sigma_{12}$\,, and 
$\tilde{\Sigma}_2$ follows by $1\leftrightarrow 2$\,; 
finally, $A-B\equiv 1-G_1G_2G_{12}^{\; 2}$ and $A+B\equiv
1+G_1G_2\Sigma_{12}^{\; 2}$\,. We recall that `products' of 
two-point functions involve integrations over intermediate 
coordinates, which we suppressed as before.    
The two-point functions are assumed to be translation invariant (bulk matter).  
  
Even for the simple bilinear interaction the effective action is quite 
complicated due to the full Hartree-Fock approximation for 
the system and environment fields. Note the 
$O(\hbar )$ and $O(\hbar^2)$ quantum corrections to the classical 
action, which appears in the first line of eq.\,(\ref{124}).     
We observe that the dressing of the two-point functions 
$G_1$,$G_2$ by a geometric series of terms involving 
themselves and $G_{12}$ disappears, as soon as the latter 
correlation function vanishes. The field equations 
resulting from the variations of the effective action are 
equally involved. In Section 4.3 we present a 
simplified version 
of them and some intriguing numerical results concerning 
the zero-dimensional limit, i.e. the quantum mechanical Hartree-Fock approximation.   
 
\subsection{The Entropy Functional}    

Knowing formally the wave functional of the system 
($\Phi_1$) coupled to the environment ($\Phi_2$), 
we presently evaluate the von\,Neumann entropy of the system in terms of the two-point functions of the previous section and according to the outline in Section 2\,. 
  
First of all, the system density functional is obtained 
from the total density functional (matrix) by tracing over the environment d.o.f., cf. eq.\,(\ref{rhosys}):
\beq 
\rho_S[\Phi_1,\Phi_1';t]=\int {\cal D}\Phi_2
\langle|\hat\rho (t) |\Phi_1',\Phi_2\rangle
=\int {\cal D}\Phi_2\;\Psi _{12}^\ast [\Phi _1',\Phi _2;t]
\Psi _{12}[\Phi _1,\Phi _2;t]
\;\;, \eeq{rhosysf}
which is a Gaussian integral again. We obtain:   
\beq
\rho_S[\Phi_1,\Phi_1';t]=   
\tilde{\Psi}_{G_1}^\ast [\Phi_1';t]
\tilde{\Psi}_{G_1}[\Phi_1;t]\;\exp\left\{
Y_1^\ast [\Phi_1';t]G_2(t)
Y_1[\Phi_1;t]\right\}
\;\;, \eeq{rhosysff}
with:  
\beq
Y_1[\Phi ;t]\equiv
\half [\Phi -\bar{\Phi}_1]\left [G_{12}(t)-i\Sigma _{12}(t)\right ]
\;\;, \eeq{113}
and where the effective Gaussian $\tilde{\Psi}_{G_1}$ here is defined as before, cf. eqs.\,(\ref{91}) and (\ref{100}), however, with the  replacements:
\beqar
N_1(t)&\longrightarrow&\tilde{N}_1(t)\equiv
N_1(t)N_{12}(t)\;\;, \nonumber \\ [1ex]
G_1^{-1}(t)&\longrightarrow&\tilde{G}_1^{-1}(t)\equiv
G_1^{-1}(t)A(t)\;\;, \nonumber \\ [1ex]
\Sigma_1(t)&\longrightarrow&\tilde{\Sigma}_1(t)
\;\;, \eeqar{replace} 
with $A$ and $\tilde\Sigma_1$ as introduced after eq.\,(\ref{124}). Note that the result of eq.\,(\ref{113}) 
has the typical form of a modified pure state density 
matrix times an exponential influence functional; both 
modifications vanish in the limit of vanishing correlations 
between system and environment.  
  
In order to evaluate the entropy, i.e. -Tr$\rho_S$ln$\rho_S$\,, we need to diagonalize the 
density matrix, such that its eigenvalues are accessible. 
This calculation was performed by functional means in 
Ref.\,\cite{I}(b).  
The result can be represented in the form: 
\beq
S_S(t)=-\int\frac{\dm^dk}{(2\pi )^d}\left\{
\mbox{ln}(1-Y_k)+\frac{Y_k}{1-Y_k}\mbox{ln}Y_k\right\}
\;\;, \eeq{Ssysk}
where $X_k$ denotes the d-dimensional Fourier transform 
of $X(x)$, and we find: 
\beqar \label{Yk}
Y_k&=&\frac{B_k}{A_k+(A_k^{\; 2}-B_k^{\; 2})^{1/2}}
\\ [1ex]
&\approx&\kappa [G_1G_2(G_{12}^{\; 2}+\Sigma_{12}^{\; 2})]_k 
\;\;, \eeqar{Ykappr}
in terms of $A$,\,$B$ as before, and where the constant 
$\kappa$ is 1/4 (1/2) in the small (large) entropy limit. 
We employed here that the products of   
translation invariant two-point functions   
involving integrations over intermediate coordinates 
become convolutions, which factorize after Fourier transformation.    
The result in eq.\,(\ref{Ssysk}) represents a sum of 
oscillator like terms, which could have been expected, cf.  
the first of Refs.\,\cite{Mark}. 
   
Several remarks are in order here: 
\begin{itemize} 
\item The results obtained are formally independent 
of the dynamics, which only enters through the actual time dependence 
of the two-point functions (the `mean fields'  $\bar\Phi_j$,\,$\bar\Pi_j$ do not contribute); however, their generality is 
restricted by the underlying ansatz of a Gaussian wave 
functional (TDHF), eq.\,(\ref{100}).  
\item For {\it vanishing correlations} between system and 
environment (independent subsystems), $G_{12}=\Sigma_{12}=0$\,, we obtain $S_S(t)=0$\,. 
\item For small widths of the Gaussians, $G_1$ or 
$G_2$ small, the entropy is small. In this case  
the system or the environment follows a {\it 
quasi-classical trajectory} in 'coordinate' (field) space; 
the widths cannot be squeezed to zero because of the 
uncertainty principle, which is incorporated TDHF.   
\end{itemize} 
  
Employing the entropy, eq.\,(\ref{Yk}), we  
define a dynamical time scale: 
\beq
\tau_{(D,equ)}^{-1}\equiv\frac{\dm}{\dm t}\mbox{ln}S_S(t)
\approx (\frac{\int\dot Y_k\mbox{ln}Y_k}
{\int Y_k\mbox{ln}Y_k}\;,\;-\frac{\int\dot Y_k/(1-Y_k)}
{\int\mbox{ln}(1-Y_k)} )
\;\;, \eeq{timesc}
which determines the decoherence time $\tau_D$ and the 
equilibration time $\tau_{equ}$ in the limits of small 
and large entropies, respectively, as indicated. 
  
The calculation of 
Ref.\,\cite{I}(b) yields as a side product the most 
probable eigenstate of the system density matrix, i.e. 
the {\it field pointer state} with the largest probability and 
the smallest (field) kinetic energy. It is  
a coherent state, i.e. Gaussian wave functional centered 
around the classical field configuration $\bar\Phi_1(x,t)$ 
with momentum $\bar\Pi_1$ and effective real width: 
\beq 
G_{eff}=\frac{4G_1}{1-G_1G_2(G_{12}^{\; 2}
-\Sigma_{12}^{\; 2})
-(G_1G_2G_{12}\Sigma_{12})^2}
\;\;, \eeq{width} 
to be compared with the bare width 4$G_1$\,, 
in the absence of correlations. The special role of 
coherent states as pointer states of the electromagnetic 
field has also be found more recently in Ref.\,\cite{ZA}, 
in a model of a dielectric medium.  

\subsection{Equations of Motion and Semiquantum Chaos}  

In order to illustrate the dynamical content of the effective 
action derived in Section 4.1\,, we consider simplified 
versions of the resulting equations of motion in this section, 
in particular also the case without environment. The full set of 
equations for arbitrary quartic interactions within and between 
system and environment can be found in Ref.\,\cite{I}(a).     
  
Presently, we make the additional assumptions:  
i) We study only the infrared limit, i.e. quantum  
mechanics of spatially homogeneous coupled fields $\Phi_1,\,\Phi_2$\,. 
ii) We neglect mean fields in the environment, 
$\bar\Phi_2,\,\bar\Pi_2\approx 0$\,, and consider the case of small  
correlations $G_{12},\,\Sigma_{12}$ between system and environment.   
  
Varying the effective action, eq.\,(\ref{124}), and 
applying the above assumptions, we find the     
set of coupled nonlinear (first order) equations of motion : 
\beqar
\frac{\delta\Gamma}{\delta\bar{\Pi}_1}=0
&\Longrightarrow&
\partial_t\bar{\Phi}_1=\bar{\Pi}_1
\;\;, \label{125} \\ [1ex] 
\frac{\delta\Gamma}{\delta\bar{\Phi}_1}=0
&\Longrightarrow&
\partial_t\bar{\Pi}_1=
-v_1^{(1)}
-\frac{\hbar}{2}v_1^{(3)}G_1
\;\;, \label{126} \\ [1ex]
\frac{\delta\Gamma}{\delta\Sigma_j}=0
&\Longrightarrow&
\partial_tG_j=4\Sigma_jG_j
\;\;, \label{127} \\ [1ex]
\frac{\delta\Gamma}{\delta\Sigma_{12}}=0
&\Longrightarrow&
\partial_tG_{12}=
-2(\Sigma_1+\Sigma_2)G_{12}
-\frac{1}{2}(G_1^{-1}+G_2^{-1})\Sigma_{12}
\;\;, \eeqar{128}
Furthermore ($j'\neq j,\,j=1,2$):
\beqar
\frac{\delta\Gamma}{\delta G_j}=0
\Longrightarrow
\partial_t\Sigma_j
=-2\Sigma_j^{\;2}
+\frac{1}{8}G_j^{-2}
-\frac{1}{2}v_j^{(2)}
-\frac{\hbar}{4}v_j^{(4)}G_j
-\mu^2G_{j'}G_{12}
\;, \label{129} \\ [1ex]
\frac{\delta\Gamma}{\delta G_{12}}=0
\Longrightarrow
\partial_t\Sigma_{12}
=
-2(\Sigma_1+\Sigma_2)\Sigma_{12}
+\frac{1}{2}(G_1^{-1}+G_2^{-1})\Sigma_{12}
+2\mu^2
\;, \eeqar{130}
where all quantities are simply functions of time; we combined the 
resulting equations in such a way that time derivatives on the r.h.s. were eliminated. Note that the terms 
$\propto\mu^2$ are due to the system-environment coupling. 

Several features of eqs.\,(\ref{125})--(\ref{130}) appear to be of a rather general nature. 
First of all, the `momenta' $\bar\Pi_1,\,\Sigma_j$ can 
be easily eliminated. The resulting second order equations are 
of the (inhomogeneous) anharmonic oscillator type. They   
potentially become unstable due to a dynamical `negative mass squared' 
term; in this case the effective potential corresponding to $\Gamma$ 
has a complicated structure allowing for tunneling processes \cite{Blum}. 

Furthermore, the coupled linear 
eqs.\,(\ref{128}),\,(\ref{130}) can be integrated
formally. We consider here only the solution of one of them 
in terms of the other:   
\beq 
G_{12}(t)=
G_{12}(0)\mbox{e}^{-2\int_{0}^t\dm t''(\Sigma_1+\Sigma_2)}
-\half\int_0^t\dm t'\;\Sigma_{12}\left (G_1^{-1}+G_2^{-1}
\right )\mbox{e}^{-2\int_{t'}^t\dm t''(\Sigma_1+\Sigma_2)} 
, \eeq{G12} 
which shows that the 
evolution of the system-environment correlations, and therewith of 
the whole dynamics, involves characteristic {\it non-Markovian}   
features. Furthermore, one may expect a   
large number of nonzero  
components in the Fourier spectra of the two-point functions of the system or environment alone.  
The functions $G_{12},\,
\Sigma_{12}$ inherit these and, therefore, the non-Markovian behavior leads to a {\it quasi stochastic} influence of the environment.   

Finally, expanding $G_{12},\,\Sigma_{12}$ for sufficiently 
short times, we explicitly solve eqs.\,(\ref{128}),\,(\ref{130}). In the limit of small but 
finite correlations we obtain: 
\beqar
G_{12}(t)&\approx&G_{12}-2[(\Sigma_1+\Sigma_2)G_{12}
+\quarter (G_1^{-1}+G_2^{-1})\Sigma_{12}]t
\nonumber \\ 
&\;&
-\half\mu^2[G_1^{-1}+G_2^{-1}+\mbox{O}(G_{12},\Sigma_{12})]
t^2
\;\;, \label{G12sol} \\ [1ex]
\Sigma_{12}(t)&\approx&\Sigma_{12}+2[\mu^2
-(\Sigma_1+\Sigma_2)\Sigma_{12}
+\quarter (G_1^{-1}+G_2^{-1})G_{12}]t
\nonumber \\ 
&\;&
-2\mu^2[\Sigma_1+\Sigma_2+\mbox{O}(G_{12},\Sigma_{12})]
t^2
\;\;, \eeqar{S12sol}
where all two-point functions on the r.h.s. here assume their initial 
value ($t=0$). Note  
how the analytical behavior changes as 
$G_{12}(0),\,\Sigma_{12}(0)\rightarrow 0$\,.           
 
This is reflected in the decoherence time $\tau_D$ 
of our simple model, 
which we calculate employing 
eqs.\,(\ref{Yk})--(\ref{timesc}) of the previous 
section and assuming small but finite $G_{12},\,\Sigma_{12}$\,:  
\beq 
\tau_D^{-1}\approx\frac{\dm}{\dm t}\mbox{ln}
[G_1G_2(G_{12}^{\;2}+\Sigma_{12}^{\;2})]_{t\rightarrow 0}
\approx 4(\Sigma_1(0)+\Sigma_2(0))
+\frac{4\mu^2\Sigma_{12}(0)}
{G_{12}(0)^{\;2}+\Sigma_{12}(0)^{\;2}}
\;. \eeq{tauD} 
In the absence of the initial correlations $G_{12}(0),\,\Sigma_{12}(0)$ the decoherence 
time vanishes instead,   
$\tau_D\sim t/2$ ($t\rightarrow 0$).     
This demonstrates the important role of correlations, 
which has been investigated in more detail 
in models of 
quantum Brownian motion (cf. Section 3), see e.g.  
Refs.\,\cite{Grab,RP}. For an environment without self-interaction at high temperature one finds  
$\tau_D\propto T$ \cite{Zu}.         

For the remaining part of this section we neglect the 
environment, in order to point out an important effect of  
the TDHF approximation employed in the derivation of the 
equations of motion (\ref{125})--(\ref{130}). Considering  
the effective action, eq.\,(\ref{124}), it has become obvious that 
this approximation includes only the O($\hbar$) and O($\hbar^2$) 
corrections to the classical action. 

Whereas the classical equations of motion 
of the system alone (1d anharmonic oscillator $\equiv$ 2 d.o.f.) can have only regular solutions according to the Poincar\'e-Bendixson theorem 
\cite{Schuster}, the potential for {\it semiquantum chaos} is obvious 
in the case of the four coupled nonlinear first order equations, 
cf. eqs.\,(\ref{125})--(\ref{127}) and (\ref{129}), representing the 
system in TDHF. A detailed study of this subset of equations has been made 
in Ref.\,\cite{Blum}. 
  
Here we represent some illustrating numerical results obtained 
by integrating the  
equations of motion forward in time. 
The initial configuration is a Gaussian wave packet with  
conserved energy $E$ (all quantities are suitably rescaled  
and presented in dimensionless form). For the chosen 
model parameters the minimum of the energy is at $E=-24.3$\,, the top 
of the hill 
separating the two wells of the classical {\it double-well 
oscillator} potential is at $E=0.0$\,.

In Fig.\,4 Poincar\'e sections for various energies are shown 
($\phi\equiv\bar\Phi (t)$, $\pi\equiv\bar\Pi (t)$). 
We observe the 
characteristic break-up of KAM tori as the energy is increased and 
the corresponding stochastic filling of the $\phi ,\,\pi$ phase space \cite{Schuster}. 
This clearly shows a transition between regular and chaotic motion, 
which can be associated with tunneling paths,  
despite the fact that the present classical model is regular 
\cite{Blum}.  
 
Furthermore, in Fig.\,5 the behavior the 
Fourier transform w.r.t. time of a diagonal 
element of the density matrix is shown. At very low energies (not 
shown) one finds a simple line spectrum involving the fundamental 
frequency $\omega_0$ (i.e. harmonic approximation of the 
potential) or a few multiples thereof (regular motion). 
At the energies shown the full trajectories are chaotic and 
produce a broadband noise spectrum. At very high energies the motion 
becomes regular again with the wave packet experiencing essentially 
only the anharmonic quartic part of the potential.   

\begin{figure}[htbp]
\centerline{
\epsfysize=7.0cm
\epsffile{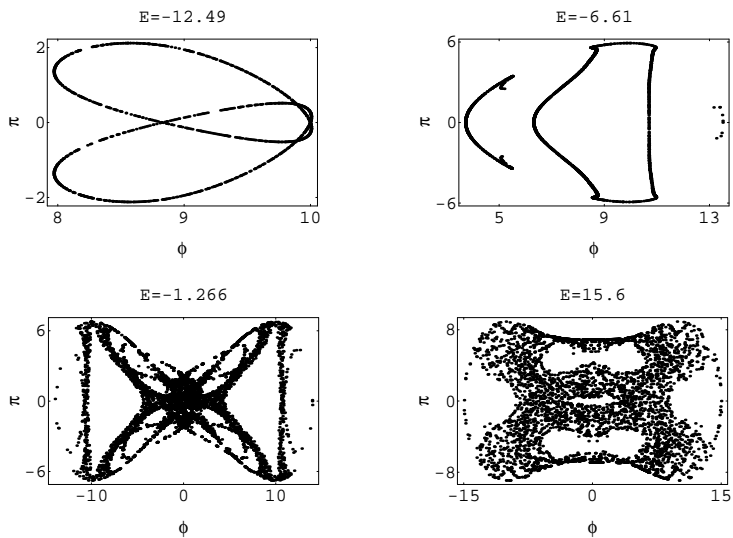}}
\caption{Poincar\'e sections for the system at various energies 
(after Ref.\,[20]) showing the transition to 
semiquantum chaos, cf. main text.
\hfill .}
\end{figure}
\begin{figure}[htbp]
\centerline{
\epsfysize=7.0cm
\epsffile{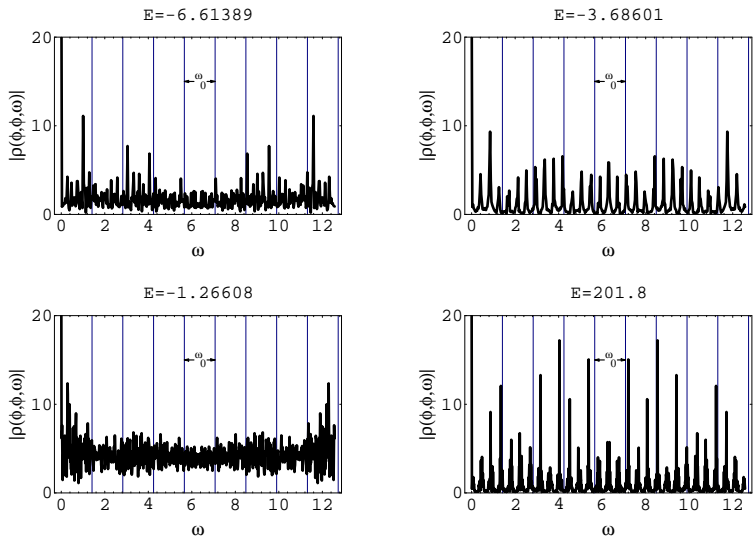}}
\caption{Fourier transform of a diagonal element of the system 
density matrix at various energies (after Ref.\,[20]) showing 
semiquantum chaos as broadband noise in the expected line spectrum 
(nonlinear resonance at E=-3.68601), cf. main text.
\hfill .}
\end{figure}
 
These results raise the question of the reliability of the 
underlying TDHF approximation or analogous semiclassical perturbative 
expansions around the bare ($\hbar =0$) classical limit. 
Generally, the 
full quantum evolution according to the Schr\"odinger equation incorporating  
an environment implies a {\it linear but 
non-Markovian master 
equation} for the system density matrix, which might rule out the possibility of `quantum chaos'. 

Furthermore,   
it has recently been shown that {\it any} variational approximation for a closed system  
is equivalent to an effective classical Hamiltonian dynamics for the 
variational parameters \cite{Cooperetal}. There, as we have seen 
in the TDHF case, 
eqs.\,(\ref{125})--(\ref{130}), the potential for {\it semiquantum chaos} arises, if the underlying classical Lagrangian involves higher than quadratic terms. 
In some examples then, the onset of semiquantum chaos 
has indeed been related to the breakdown of semiclassical approximations \cite{Cooperetal}, even if they do not require weak coupling. 
 
However, the important question remains \cite{Blum}, 
whether there are situations involving (infinitely) many d.o.f., where such approximations could reliably represent the highly complex, even if overall 
quasi-periodic     
quantum evolution of the coupled system+environment (fields). Such cases have recently been found  
in solid state devices. Nevertheless, the real time evolution of fields which are relevant for high-energy experiments still needs more study, in order to 
understand decoherence processes, entropy production, and thermalization quantitatively.             
  
\section{Conclusions} 
We have seen in Sections 1,\,2 that entropy production in a  
quantum system can be 
understood dynamically on the basis of von\,Neumann's definition 
of the entropy in terms of its density operator, if and only if 
the system is coupled to an environment of unobserved or even unobservable 
degrees of freedom. It induces the 
necessary quantum decoherence \cite{Zu}.  
  
The example of a nonrelativistic quark in Brownian motion in a 
simple model of a gluonic environment has been presented (Section 3). 
It has been pointed out that an environment with a 
spectral density distribution which is enhanced in the infrared, 
as compared to the frequently studied `Ohmic', phonon, or photon  
environments \cite{Grab}, may lead to interesting squeezing, 
stabilizing, or enhanced spreading of the wave packet effects, 
which affect the rate of entropy production. 
 
In order to extent the decoherence approach to field theory,    
a system and environment consisting of two distinct scalar 
fields have been studied in the time dependent Hartree-Fock 
approximation in Section 4 \cite{I}. We derived the effective action including 
up to O($\hbar^2$) corrections, the equations of motion, and the entropy 
functional and discussed their general properties. 

Particular attention  
has been paid to the fact that the approximate semiclassical dynamics may 
be intrinsically unstable and leading to deterministic semiquantum chaos.  
It might signal a breakdown of any such approximation scheme for genuinely 
nonlinear interacting field theories. Whether under these circumstances 
the initial decoherence, entropy generation, and possibly thermalization 
in high-energy hadronic or nuclear collisions can be reliably calculated, is an interesting question. 
The decoherence effects  
due to initial bremsstrahlung mentioned in Section 2 are presently studied.             

\section*{Acknowledgements}  
I thank the members of the   
Instituto de F\'{\i}sica (UFRJ) for their great hospitality 
and particularly T. Kodama for many stimulating 
discussions and his support. Discussions with C. E. Aguiar, J. P. Paz, J. Rafelski, S. Rugh and 
W. Zurek 
are gratefully acknowledged.          
This research was supported in part by US-Department of Energy
under Grant No. DE-FG03-95ER40937, by NSF under grant INT-9602920,
and by Brazil-PRONEX-41.96.0886.00\,.


\section*{References} 
\begin{thebibliography}{99}

\bibitem{Zu} W. H. Zurek, Phys. Today {\bf 44}, No. 10, 36 (1991).

\bibitem{Om} R. Omn\`es, \rmp{{\bf 64}}, 339 (1992).

\bibitem{Ze} H. D. Zeh, \pl{A\,{\bf 172}}, 189 (1993).

\bibitem{GH} M. Gell-Mann and J. B. Hartle, \prd{\,{\bf 47}}, 3345 
(1993); \\
in {\it Complexity, Entropy and the Physics of Information}, 
ed. W. H. Zurek (Addison-Wesley, Redwood City, CA., 1990). 

\bibitem{Ellis} J. Ellis, N. E. Mavromatos and D. V. Nanopoulos, 
Mod. Phys. Lett. A\,{\bf 12}, 1759 (1997);  
CERN-TH.7000/93 -- hep-th/9311148; \\ \pl{B\,{\bf 293}}, 37 (1992). 

\bibitem{I} H.-Th. Elze, \np{B\,{\bf 436}}, 213 (1995) (a); \\ 
\pl{B\,{\bf 369}}, 295 (1996) (b); \\ 
in {\it Quantum Infrared 
Physics}, eds. H. M. Fried and B. M\"uller (World 
Scientific, Singapore, 1995) -- hep-ph/9407377 (c).  

\bibitem{QM} P. Braun-Munzinger et al., eds., {\it 
Quark Matter '96} (North-Holland, Amsterdam, 1996).  

\bibitem{Feyn} R. P. Feynman, {\it Statistical Mechanics} (Benjamin, Reading, Mass., 1974).

\bibitem{EH} H.-Th. Elze and U. Heinz, \prep{{\bf 183}}, 81 (1989).

\bibitem{EC} H.-Th. Elze and P. A. Carruthers, in 
{\it Particle Production in Highly Excited Matter}, eds. H. H. Gutbrod, 
J. Letessier and J. Rafelski, NATO ASI series B:
Vol. 303 (Plenum, New York, 1994) -- hep-ph/9409248.

\bibitem{Te} M. Tegmark, Found. Phys. Lett. {\bf 6}, 571 (1993).
 
\bibitem{Grab} H. Grabert, P. Schramm and G.-L. Ingold, 
\prep{{\bf 168}}, 115 (1988).

\bibitem{Schuster} H. G. Schuster, {\it Deterministic Chaos}, 2nd ed. (VCH Publishers, Weinheim and New York, 1989).

\bibitem{ZHP} W. H. Zurek, S. Habib and J. P. Paz, \prl{{\bf 70}}, 
1187 (1993).

\bibitem{Mark} M. Srednicki, \prl{{\bf 71}}, 666 (1993); \\
D. Kabat, \np{B\,{\bf 453}}, 281 (1995); \\
E. Benedict and S.-Y. Pi, \ap{{\bf 245}}, 209 (1996). 

\bibitem{LM} F. Lombardo and F. D. Mazzitelli, \prd{\,{\bf 53}},  
2001 (1996). 
 
\bibitem{Jackiw} R. Jackiw and A. Kerman, \pl{{\bf 71}\,A}, 158 (1979).

\bibitem{Corn} J. M. Cornwall, R. Jackiw and E. Tomboulis, 
\prd{\,{\bf 10}}, 2428 (1974).

\bibitem{ZA} J. R. Anglin and W. H. Zurek, \prd{\,{\bf 53}}, 7327 (1996).  

\bibitem{Blum} Th. C. Blum and H.-Th. Elze, \pre{\,{\bf 53}}, 3123 (1996).  

\bibitem{RP} L. D. Romero and J. P. Paz, quant-ph/9612036.  

\bibitem{Cooperetal} F. Cooper, J. Dawson, S. Habib and R. D. Ryne, 
LA-UR-96-3335 -- quant-ph/9610013.   

\end{thebibliography}
\end{document}